\documentclass{ptephy_v1}

\usepackage{amsmath}
\usepackage{graphics}
\usepackage{url}

\usepackage{color}

\allowdisplaybreaks

\def\edth{{\rlap{$\partial$}\raise0.3em\hbox{$-$}}}
\newcommand{\bea}{\begin{eqnarray}}
\newcommand{\eea}{\end{eqnarray}}

\begin{document}

\title{Pre-DECIGO can get the smoking gun to decide the astrophysical or cosmological origin of GW150914-like binary black holes}


\author{\name{Takashi Nakamura}{1}, 
\name{Masaki Ando}{2,3,5},
\name{Tomoya Kinugawa}{4},
\name{Hiroyuki Nakano}{1},
\name{Kazunari Eda}{2,5},
\name{Shuichi Sato}{6},
\name{Mitsuru Musha}{7},
\name{Tomotada Akutsu}{3},
\name{Takahiro Tanaka}{1,8},
\name{Naoki Seto}{1},
\name{Nobuyuki Kanda}{9},
\name{Yousuke Itoh}{5}}

\address{${}^1$\affil{1}{Department of Physics, Kyoto University, Kyoto 606-8502, Japan}
\\
${}^2$\affil{2}{Department of Physics, The University of Tokyo, Tokyo 113-0033, Japan}
\\
${}^3$\affil{3}{National Astronomical Observatory of Japan, Tokyo 181-8588, Japan}
\\
${}^4$\affil{4}{Institute for Cosmic Ray Research, The University of Tokyo,
Chiba 277-8582, Japan}
\\
${}^5$\affil{5}{Research Center for the Early Universe (RESCEU), The University of Tokyo, Tokyo 113-0033, Japan}
\\
${}^6$\affil{6}{Department of Advanced Sciences, Hosei University, Tokyo 184-8584 Japan}
\\
${}^7$\affil{7}{Institute for Laser Science, University of Electro-Communications, Tokyo 182-8585, Japan}
\\
${}^8$\affil{8}{Yukawa Institute for Theoretical Physics, Kyoto University, Kyoto 606-8502, Japan}
\\
${}^9$\affil{9}{Department of Physics, Osaka City University, Osaka 558-8585, Japan}
}

\begin{abstract}
Pre-DECIGO (DECihertz laser Interferometer Gravitational wave Observatory)
consists of three spacecraft arranged in an equilateral triangle 
with 100\,km arm lengths orbiting 2000\,km
above the surface of the earth.
It is hoped that the launch date will be in the late 2020s.

Pre-DECIGO has one clear target: binary black holes (BBHs)
like GW150914 and GW151226.
Pre-DECIGO can detect $\sim 30M_\odot$--$30M_\odot$ BBH mergers
like GW150914 up to redshift $z\sim 30$. 
The cumulative event rate is $\sim 1.8\times 10^{5}\,{\rm events~yr^{-1}}$ 
in the Pop III origin model of BBHs like GW150914,
and it saturates at $z\sim 10$, while in the primordial BBH (PBBH) model,
the cumulative event rate is $ \sim 3\times 10^{4}\,{\rm events~ yr^{-1}}$ at $z=30$
even if only $0.1\%$ of the dark matter consists of PBHs,
and it is still increasing at $z=30$. In the Pop I/II model of GW150914-like BBHs,
the cumulative event rate is 
$(3$--$10)\times10^{5}\,{\rm events~ yr^{-1}}$ and it saturates at $z \sim 6$.
We present the requirements on orbit accuracy,
drag-free techniques, laser power, frequency stability,
and the interferometer test mass.
For BBHs like GW150914 at 1\,Gpc ($z\sim 0.2$),
SNR~$\sim 90$ is achieved with the definition of Pre-DECIGO
in $0.01$--$100$\,Hz band. Since for $z\gg 1$
the characteristic strain amplitude $h_c$
for a fixed frequency band weakly depends on $z$ as $z^{-1/6}$,
$\sim 10\%$ of BBHs near face-on have SNR~$> 5 \ (7)$ even at $z\sim 30 \ (10)$.
Pre-DECIGO can measure the mass spectrum and the $z$-dependence of the merger rate
to distinguish various models of BBHs like GW150914, such as Pop III BBH,
Pop II BBH and PBBH scenarios.  

Pre-DECIGO can also predict the direction of BBHs at $z = 0.1$
with an accuracy of $\sim 0.3\,\deg^2$
and a merging time accuracy of $\sim 1\,$s
at about a day before the merger so that 
ground-based GW detectors further developed at that time
as well as electromagnetic follow-up observations can prepare
for the detection of merger in advance like a solar eclipse.
For intermediate mass BBHs such as $\sim 640 M_\odot$--$640 M_\odot$
at a large redshift $z > 10$, the quasinormal mode frequency after the merger
can be within the Pre-DECIGO band so that the ringing tail can also be detectable
to confirm the Einstein theory of general relativity with SNR~$\sim 35$.
\end{abstract}

\subjectindex{E02, E31, E01, F31}

\maketitle

\section{Introduction}

The first direct detection of a gravitational wave (GW) has been done by O1 of
aLIGO~\cite{Abbott:2016blz}.
The event called GW150914 was a binary black hole (BBH) of mass
$\sim 30 M_\odot$--$30 M_\odot$.
Such high mass black hole (BH) candidates had not been confirmed
although several suggestions 
existed~\cite{Kalogera:2006uj,Dominik:2012kk,Kinugawa:2014zha,Kinugawa:2015nla,
Spera:2015vkd,Amaro-Seoane:2015umi,Mandel:2015qlu,Marchant:2016wow,
Belczynski:2016obo}
so that its origin is not known at present,
while for the merger of neutron star (NS) binary
there exists several systems
with merger time less than the age of the universe
so that the event rate has been estimated~\cite{Kim:2002uw,Kalogera:2003tn,
Kim:2013tca},
and this was the most plausible GW source for the first direct detection of GWs
before GW150914~\cite{Abadie:2010cf}.
For BBHs, no observation of electromagnetic counterparts exists so that
analyzing theoretical models is the only way to provide methods
to study their properties.
One method is population synthesis, in which the Monte Carlo simulations
have been performed to evolve binaries 
starting from binary ZAMS (Zero Age Main Sequence) stars.
The mass of the observed BH candidate in X-ray binaries is at most $\sim 15 M_\odot$,
which is about half of the mass of GW150914.
This suggests that the progenitors of GW150914 are low-metal stars
with little or no mass loss such as Pop II or Pop III
stars~\cite{Baraffe:2000dp,Inayoshi:2013rfa}.
In particular, the predicted mass of Pop III BBHs
by Kinugawa et al.~\cite{Kinugawa:2014zha}
(see also Refs.~\cite{Kinugawa:2015nla,Kinugawa:2016mfs,Kinugawa:2016ect})
agrees astonishingly well with GW150914~\cite{TheLIGOScientific:2016htt}.
However, a single event is not enough to restrict the origin of
BBHs like GW150914, 
around $6$--$8$ of which will be found in O2.
The cumulative chirp mass and the total mass as well as the distribution of
the spin parameter $a/M$ of the merged BH will help to distinguish
the plausible model among the various population synthesis models.
Here, in the population synthesis Monte Carlo simulations
of Pop II and III stars, there are so many unknown functions
and parameters such as initial mass function, initial eccentric distribution function,
the distribution of initial separation, the distribution of mass ratio,
the Roche lobe overflow parameters, the common envelope parameters and so on,
so that the observed cumulative chirp mass and the total mass as well as
the distribution of the spin parameter will only give constraints
among these undetermined functions and parameters.
 
There are other completely different formation scenarios of 
BBHs like GW150914,
such as primordial BBH (PBBH)~\cite{Sasaki:2016jop}~\footnote{
They simply applied the method in Ref.~\cite{Nakamura:1997sm}
for the case with mass from $0.5M_\odot$ to $30M_\odot$.}
and the three-body dynamical formation model~\cite{Amaro-Seoane:2015umi}.
In particular, in the PBBH model the mass spectrum primarily reflects the primordial
density perturbation spectrum.
This means that it is difficult to distinguish models only
by observations of small--$z$ BBHs like GW150914
since the average detectable range of GW150914 with the design sensitivity
of aLIGO is at most $z \sim 0.3$ whose luminosity distance is $\sim 1.5\,{\rm Gpc}$. 

It is highly requested to observe GW150914-like events for larger $z$ to distinguish
various models. For this purpose, DECIGO (DECiherz laser Interferometer
Gravitational wave Observatory) is suitable although it was originally proposed
by Seto, Kawamura, and Nakamura~\cite{Seto:2001qf}
to measure the acceleration of the universe through GWs from binary NS--NS
at $z\sim 1$ and observe the chirp signal $1$--$10$ years
before the final merger.
The GW frequency from such NS-NS binaries is $\sim 0.1$\,Hz ($=$decihertz),
which is the origin of the name of DECIGO.~\footnote{
Another origin of the name is ``DECIde and GO project.''}
DECIGO consists of at least a triangle shape three spacecraft
in a heliocentric orbit
with a 1000\,km Fabry--P\'{e}rot laser interferometer to particularly observe
the GW background (GWB) from the inflation at frequency $0.1$\,Hz~\cite{Kawamura:2006up}.
Pre-DECIGO~\footnote{
Pre-DECIGO was initially the pathfinder for DECIGO.
That is, Pre-DECIGO was a smaller version of DECIGO,
and the design of Pre-DECIGO had not been defined
including the selection of possible targets,
while DECIGO has a definite design and clear targets. 
In particular, the sensitivity of Ultimate-DECIGO 
is limited only by the uncertainty principle so that it can detect
the inflation origin background GWs
even if $\Omega_{\rm GW}=10^{-20}$.}
is a smaller DECIGO which consists of
three spacecraft arranged in an equilateral triangle 
with 100\,km arm length orbiting 2000\,km
above the surface of the earth.
The orbit of Pre-DECIGO is geocentric and
is different from DECIGO whose orbit is heliocentric.
We are hoping that the launch date will be in the late 2020s.
In this paper, we show that Pre-DECIGO can detect events like GW150914
up to $z\sim 30$ where the cumulative event rate is
$\sim 1.8\times 10^{5}\,{\rm events~yr^{-1}}$ in the Pop III origin model of GW150914, while in the PBBH model, it is $\sim 3\times 10^{4}\,{\rm events~yr^{-1}}$
even if only 0.1\% of the dark matter consists of primordial BHs (PBHs).

This paper is organized as follows: In Sect. 2,
the requirements on the orbit accuracy, drag-free techniques,
laser power, frequency stability,
and interferometer test mass will be shown.
In Sect. 3, we show that Pre-DECIGO can measure the mass spectrum and
the $z$-dependence of the merger rate up to $z\sim 30$ to distinguish various models
such as Pop III BBH, Pop II BBH, PBBH, three-body dynamical formation models,
and so on.
For small $z = 0.1$, Pre-DECIGO can also predict the direction of BBHs
with accuracy $\sim 0.3\,\deg^2$, and the merger time with accuracy $\sim 1$\,s
at about a day before the merger
so that the Einstein Telescope (ET)~\cite{Punturo:2010zz}
and the enhanced version of aLIGO as well as electromagnetic
follow-up observations can prepare for the detection of the merger in advance,
like a solar eclipse. For large $z > 10$ the quasinormal mode
(QNM or ringing tail) frequency after the merger
can be within the Pre-DECIGO band so that the ringing tail can also be detectable
to confirm or refute the Einstein theory of general relativity (GR) with 
signal-to-noise ratio (SNR) $\sim 35$
for intermediate mass BBHs such as $\sim 640 M_\odot$--$640 M_\odot$.
Section 4 is devoted to discussions.

\section{Design of Pre-DECIGO}

\begin{figure}[!htb]
\begin{center}
\includegraphics[angle=270,scale=0.4,clip=true,trim=70 0 70 0]{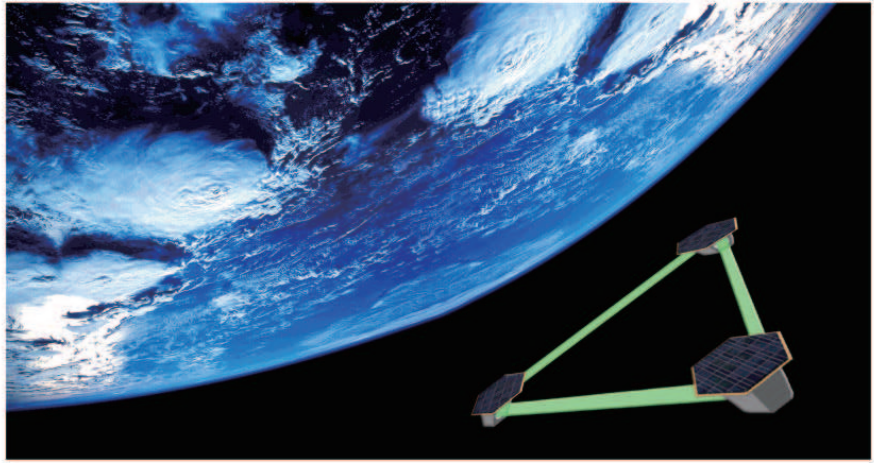}
\end{center}
\caption{Image of Pre-DECIGO which is a smaller DECIGO consisting of
three spacecraft arranged in an equilateral triangle 
with 100\,km arm lengths orbiting 2000\,km
above the surface of the earth. } 
\label{BBHDECIGOimage}
\end{figure}

Pre-DECIGO is a space-borne GW antenna with 100\,km arm lengths. 
It consists of three spacecraft separated by 100\,km
(Fig.~\ref{BBHDECIGOimage}).
Test-mass mirrors in each spacecraft have masses of 30\,kg and diameters of 30\,cm.
With these mirrors, a triangle-shaped laser interferometer unit is
formed by three 100\,km Fabry--P\'{e}rot cavities.
The finesse of the cavities is 100,
resulting in a cavity cut-off frequency around 20\,Hz.
As a laser source, frequency-doubled Yb:fiber DFB laser
with a wavelength of 515\,nm and an 
output power of 1\,W is used. 
The frequency of the laser source is pre-stabilized
in reference to the saturated absorption 
of iodine molecules, and also stabilized 
by the common-mode signal of the 100\,km arm cavities.
So as to avoid the external force fluctuation on the test mass caused by the
spacecraft motion, the test masses are kept untouched inside the spacecraft.
In addition, the spacecraft is drag-free controlled;
the displacement and attitude of the  
spacecraft are controlled by using the test masses inside it as references.

\begin{figure}[!b]  
\begin{center}
\includegraphics[width=0.9\textwidth,clip=true]{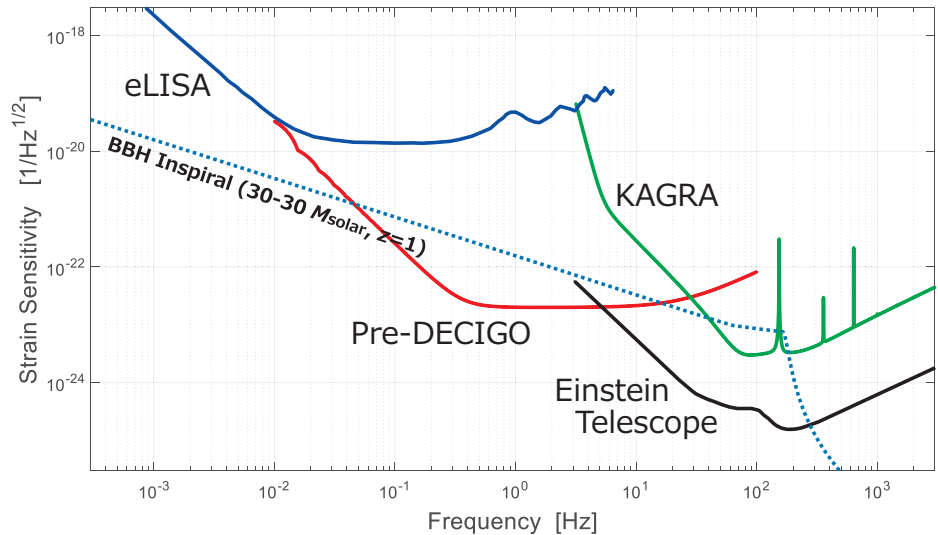}
\end{center}
 \caption{Strain sensitivity of Pre-DECIGO.
Sensitivity curves for the second-generation terrestrial
GW antenna (KAGRA~\cite{KAGRA:1999}; 
sensitivity curve at
http://gwcenter.icrr.u-tokyo.ac.jp/en/researcher/parameter),
third-generation antenna (ET~\cite{ET:2011}), and 
space antenna (eLISA~\cite{eLISA:2016}) are shown together for references.
The dashed curve shows the signal amplitude from BBH merger
with masses of $30 M_\odot$ at a distance of $z=1$. }
\label{Fig:sens}
\end{figure}  

\begin{figure}[!t]  
\begin{center}
\includegraphics[width=0.9\textwidth,clip=true]{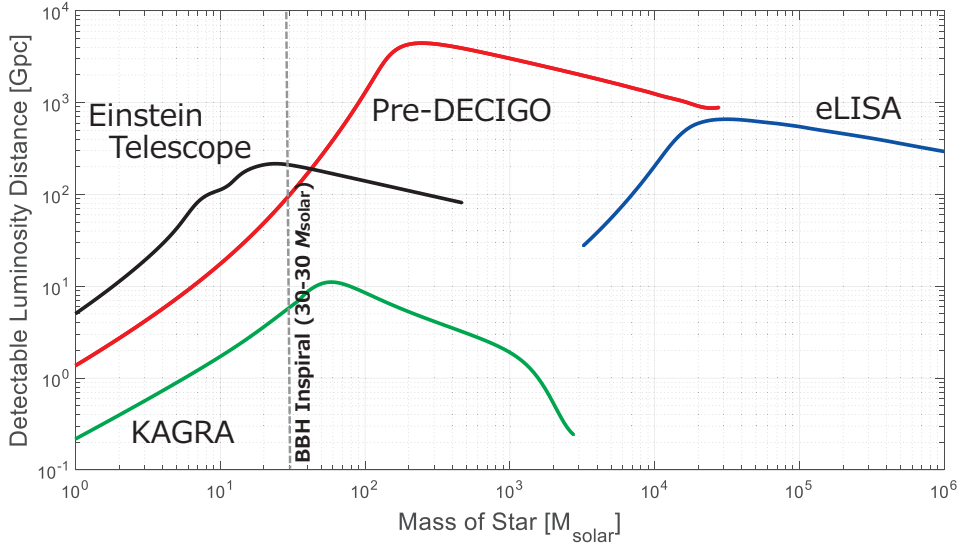}
\end{center}
 \caption{Observable range for inspiral and mergers of BBHs. Here, we assume optimal direction 
and polarization of the source, and a detection SNR threshold of 8.}
\label{Fig:range}
\end{figure}

The target sensitivity of Pre-DECIGO is $2\times 10^{-23}\, {\rm Hz^{-1/2}}$ in strain
in the current design (Fig.~\ref{Fig:sens}).
It is fundamentally limited by the optical quantum noise of the interferometer: 
laser shot noise and radiation pressure noise in high and low frequency bands, respectively.
The external force noise on the test-mass mirrors is also critical;
the requirement is $1 \times 10^{-16}\, {\rm N/Hz^{1/2}}$.
With this sensitivity, mergers of BBHs at $z=10$ will be within the
observable range of Pre-DECIGO, assuming optimal direction 
and polarization of the source, and detection SNR of 8
(Fig.~\ref{Fig:range}).

The candidate orbit of Pre-DECIGO is record-disk (or cartwheel) orbit around the earth.
The reference orbit, the orbit of the center of the mass of the three spacecraft, is 
a Sun-synchronized dusk-dawn circular orbit with an altitude of 2000\,km. 
Each spacecraft has a slightly eccentric orbit around this reference orbit so as 
to minimize the natural fluctuation in the relative distance between the spacecraft 
during the orbital motion.
Moreover, there will be no eclipse in these spacecraft, which is beneficial to avoid
thermal shock and drift in the spacecraft.
The formation flight of the three spacecraft is realized by continuous feedback control.
The laser interferometers measure the cavity length changes, which are fed back to the 
positions of the test-mass mirrors.
Since the spacecraft follows the test-mass position inside it by drag-free control,
the 100\,km triangular formation flight is realized.
The orbital period of the formation flight interferometer unit around the earth is 124\,min.
With this orbital motion and the earth's annual orbital motion around the sun,
the antenna pattern of Pre-DECIGO to observe GWs will change in time.
Parameter estimation accuracy for the GW sources, such as sky localization,
is improved because of this selection of the orbit.

\section{Physical, astronomical and cosmological targets of Pre-DECIGO}

The guaranteed science target of Pre-DECIGO is the pre-merger phase of
BBHs like GW150914.
For this purpose, we rewrite Eqs.~(3)--(5) in Ref.~\cite{Seto:2001qf}
expressing the time before the final merge $t_c$,
the number of cycle $N_c$, and the characteristic amplitude $h_c$ as
\begin{eqnarray}
 t_c &=&  1.33\times 10^6\,{\rm s}\,(1+z)^{-5/3} \left(\frac{M_1}{30M_\odot}\right)^{-1}
 \left(\frac{M_2}{30M_\odot}\right)^{-1}  \left(\frac{M_t}{60M_\odot}\right)^{1/3}
 \left(\frac{\nu}{0.1\,{\rm Hz}}\right)^{-8/3} \,, \\
 N_c &=& 1.00\times 10^5 (1+z)^{-5/3} \left(\frac{M_1}{30M_\odot}\right)^{-1}
 \left(\frac{M_2}{30M_\odot}\right)^{-1} \left(\frac{M_t}{60M_\odot}\right)^{1/3}
 \left(\frac{\nu}{0.1\,{\rm Hz}}\right)^{-5/3} \,, \\
 h_c &=& 1.89\times 10^{-21}(1+z)^{5/6} \left(\frac{M_c}{26.1M_\odot}\right)^{5/6}
 \left(\frac{\nu}{0.1\,{\rm Hz}}\right)^{-1/6} \left(\frac{d_L(z)}{1\,{\rm Gpc}}\right)^{-1} \,, 
\end{eqnarray}
where $z,\,M_1,\,M_2,\,M_t,\,M_c,\,\nu$ and $d_L(z)$ are the cosmological redshift,
the mass of each BH, the total mass, the chirp mass defined
by $(M_1M_2)^{3/5}/(M_1+M_2)^{1/5}$,
the frequency of the emitted GW assuming a circular orbit, 
and the luminosity distance, respectively.
One might think that it is difficult to determine $z$ since,
for fixed physical masses, we have $h_c\propto z^{-1/6}$
for $z \gg 1$, where $d_L(z)$ is almost proportional to $z$.
However from the phase evolution of GW,
we will get the redshifted chirp mass $(1+z)M_c$
so that from $h_c$ we will get $d_L(z)$ accurately.
Then, assuming the standard $\Lambda$CDM cosmological model value of $d_L(z)$,
we can determine the redshift $z$.
 
According to Dalal et al.~\cite{Dalal:2006qt},
the maximum and the average SNR~\footnote{
In practice, we will need some detailed study on the averaged SNR
for long-lived GW signals as discussed in Ref.~\cite{Vallisneri:2012np},
though we are using Eq.~\eqref{eq:aveSNR} as the averaged SNR
for convenience in this paper.}
are given by
\begin{eqnarray}
 {\rm SNR_{max}} &=& 4\frac{A}{d_L(z)}\sqrt{I} \,, \\
 {\rm SNR_{ave}} &=& 0.4~{\rm SNR_{max}} \,,
\label{eq:aveSNR}
\end{eqnarray}
where
\begin{eqnarray}
 I &=& \int _{\rm f_{min}}^{\rm f_{max}}\frac{f^{-7/3}}{S_n(f)}df \,,\\
 A &=& \sqrt{\frac{5}{96}}\pi^{-2/3}\left(\frac{(1+z)GM_c}{c^3}\right)^{5/6} c \,.
\end{eqnarray}
We obtain for a BBH of mass $30M_\odot$--$30M_\odot$ at 1\,Gpc ($z\sim 0.2$)
${\rm SNR\sim 90}$ in the Pre-DECIGO band of $0.01$--$100$\,Hz.
Since the characteristic strain amplitude weakly depends on the redshift $z$
as $(1+z)^{-1/6}$ for $z\gg 1$, near face-on $\sim $10\% of BBHs have
${\rm SNR}\sim 5$ even at $z\sim 30$.

\subsection{Pop III model}

\begin{figure}[!t]
\begin{center}
\includegraphics[width=0.49\textwidth,clip=true]{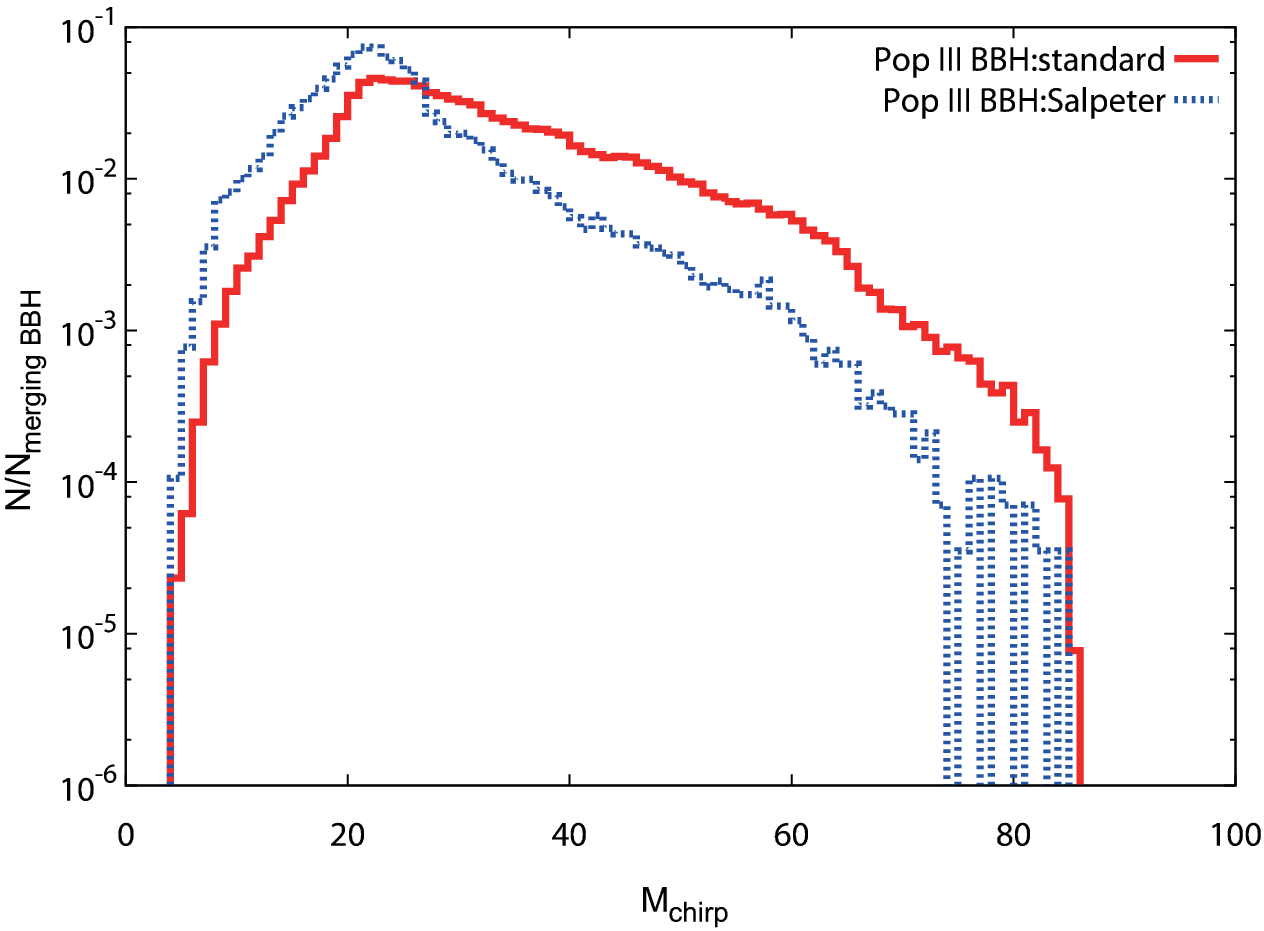}
\includegraphics[width=0.49\textwidth,clip=true]{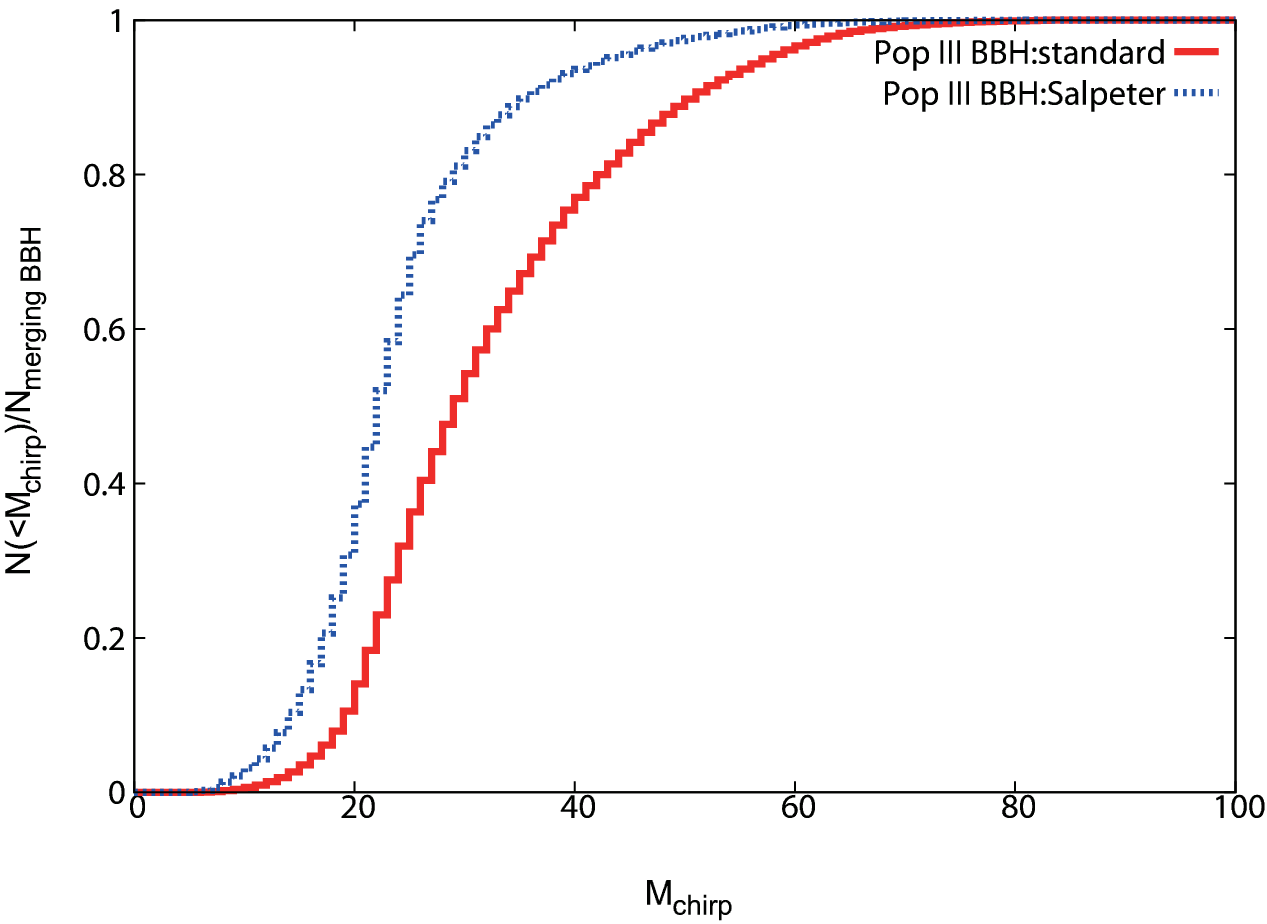}
\end{center}
 \caption{The chirp mass distributions of Pop III BBHs (left) 
 and the cumulative chirp mass distribution of Pop III BBHs (right)
 for the standard model and the Salpeter IMF model.}
\label{fig:PopIIImass}
\end{figure}

Kinugawa et al.~\cite{Kinugawa:2014zha} showed that Pop III 
binaries generally become BBHs whose mass has
a peak at $30M_\odot$--$30M_\odot$.
Figure~\ref{fig:PopIIImass} shows the Pop III BBH chirp mass distribution
and the cumulative chirp mass distribution of
the standard model defined in Kinugawa et al.~\cite{Kinugawa:2015nla}
where the initial mass range is $10M_\odot<M<140M_\odot$
with flat initial mass function (IMF).
We also give a plot for the model with the Salpeter IMF
(see Ref.~\cite{Kinugawa:2015nla} for details about the other model parameters).
The main reasons for this peak of chirp mass distribution are as follows:
First, Pop III stars tend to be born as massive stars whose typical mass is
$10$--$140M_\odot$,~\footnote{A Pop III star with mass larger
than $140M_\odot$ explodes as a pair instability supernova
or fragments.}
i.e., more massive than Pop I and II stars
because of the larger Jeans mass of the initial gas cloud
due to the lack of coolant and the weaker radiative feedback
due to the lack of absorption of photons by the metal.
Furthermore, since the stellar wind mass loss of Pop III stars
is not effective, the stellar evolution makes them  
more massive compared with Pop I and II stars.
Second, Pop III star evolution is qualitatively different from
that of Pop I and II.
Pop I and II stars evolve as a red giant at the late phase.
On the other hand, Pop III stars whose masses are less than $50 M_\odot$
evolve as a blue giant.
If a star in a binary becomes a red giant,
the mass transfer from such a giant star
to the companion star often becomes unstable and
they form a common envelope phase, during which
the companion star plunges into the envelope of the giant star
and spirals in. In this phase, the friction between the envelope and
the companion star yields a loss of orbital angular momentum
to decrease the binary separation, while the envelope will be evaporated
by the energy liberated through the friction. Consequently,
the binary becomes either a close binary that consists of the core of the giant
or a single star absorbing the companion star during the common envelope phase.
Therefore, all Pop I and II binaries initially close enough to
become a close compact binary that merges within the age of the universe,
evolve via a common envelope phase.

However, the mass transfer of a blue giant Pop III star is so stable
that Pop III stars whose mass is less than $50 M_\odot$ do not experience
a violent mass loss process such as the common envelope phase.
They evolve via stable mass transfer phases and their mass loss is smaller than
that in the evolution via a common envelope phase.
They tend to lose only a minor fraction ($1/10$--$1/3$)
of their initial mass so that they tend to end up with $20$--$30M_\odot$ BHs,
while Pop III stars whose mass is larger than $50M_\odot$ are also likely
to form a common envelope phase, and they lose $1/2$--$2/3$ of their mass
so that they tend to end up with $25$--$30M_\odot$ BHs, too.
Therefore, the peak of the Pop III BBH chirp mass distribution at
$20$--$30M_\odot$ can be understood as a natural consequence
of the evolution of Pop III stars
(see Fig.~\ref{fig:PopIIImass}).

\begin{figure}[!t]
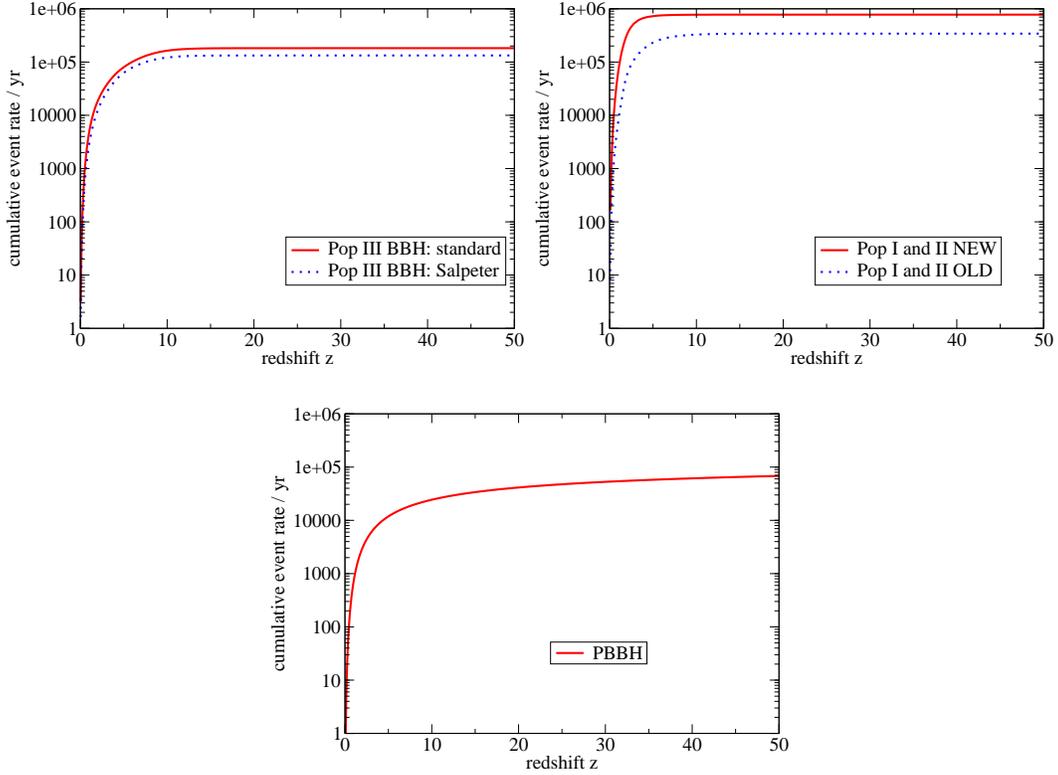

\begin{center}
 \includegraphics[width=0.45\textwidth,clip=true]{./PopIII.eps}
 \includegraphics[width=0.45\textwidth,clip=true]{./PopIandII.eps}

\vspace{5mm}

 \includegraphics[width=0.45\textwidth,clip=true]{./PBBH.eps}
\end{center}
 \caption{
Top left:
the cumulative merger rate of Pop III BBHs for the standard model
and the Salpeter IMF model.
Top right: the cumulative merger rate of Pop I and II BBHs for Belczynski's
 NEW model and OLD model~\cite{Belczynski:2016obo}.
Bottom: the cumulative event rate per year of primordial binary BH (PBBH) merger
as a function of $z$.
This corresponds to the fraction $f=0.001$; that is, PBH contributes only 0.1\%
of the dark matter. Note that the rate is still increasing even at $z=30$.}
\label{fig:CER}
\end{figure}

Kinugawa et al.~\cite{Kinugawa:2014zha,Kinugawa:2015nla}
showed the merger rate of Pop III BBHs
using their Pop III binary population synthesis result
and the Pop III star formation rate (SFR) of de Souza et al.~\cite{deSouza:2011ea}.
The Pop III BBH merger rate density of the standard (Salpeter IMF) model
at the present day is
$25 \,(13) \,
{\rm events~yr^{-1}\,Gpc^{-3}}~({\rm SFR_p}
/(10^{-2.5}M_\odot\,{\rm yr^{-1}\,Mpc^{-3}}))
\cdot (\rm [f_b/(1+f_b)]/0.33)$ 
where ${\rm SFR_{p}}$ and ${\rm f_b}$ are the peak value of the Pop III SFR and the binary fraction, respectively.
The top left panel of Fig.~\ref{fig:CER}
shows the cumulative merger rate of Pop III BBHs
for the standard model and the Salpeter IMF model.
The peak of Pop III BBH merger rate density is at $z\sim8$
because the Pop III star formation peak is at $z\sim9$~\cite{deSouza:2011ea}.
Thus, the cumulative merger rate is saturated at $z\sim10$.

\subsection{PopII/I model}

It is usually believed that Pop I and II stars are lighter than Pop III stars
because the minimum Jeans mass of Pop I and II stars is lighter than
that of Pop III, which is caused by the difference in cooling mechanism; 
that is, metal cooling and dust cooling
are at work for Pop I and II, while only hydrogen atom cooling
and molecular cooling apply for Pop III~\cite{Omukai:2005hv}. 
Although the Pop II IMF is not known well, the Pop I IMF is known rather well
by observations so that the typical mass of Pop I is
$\leq1M_\odot$~\cite{Salpeter:1955it,Kroupa:2000iv}.
Furthermore, Pop I and II stars lose mass via stellar wind due to the absorption 
of photons by the metal.
Therefore, the Pop I and II compact binaries tend to be lighter than
those of Pop III.

\begin{figure}[!t]
\begin{center}
\includegraphics[width=0.49\textwidth,clip=true]{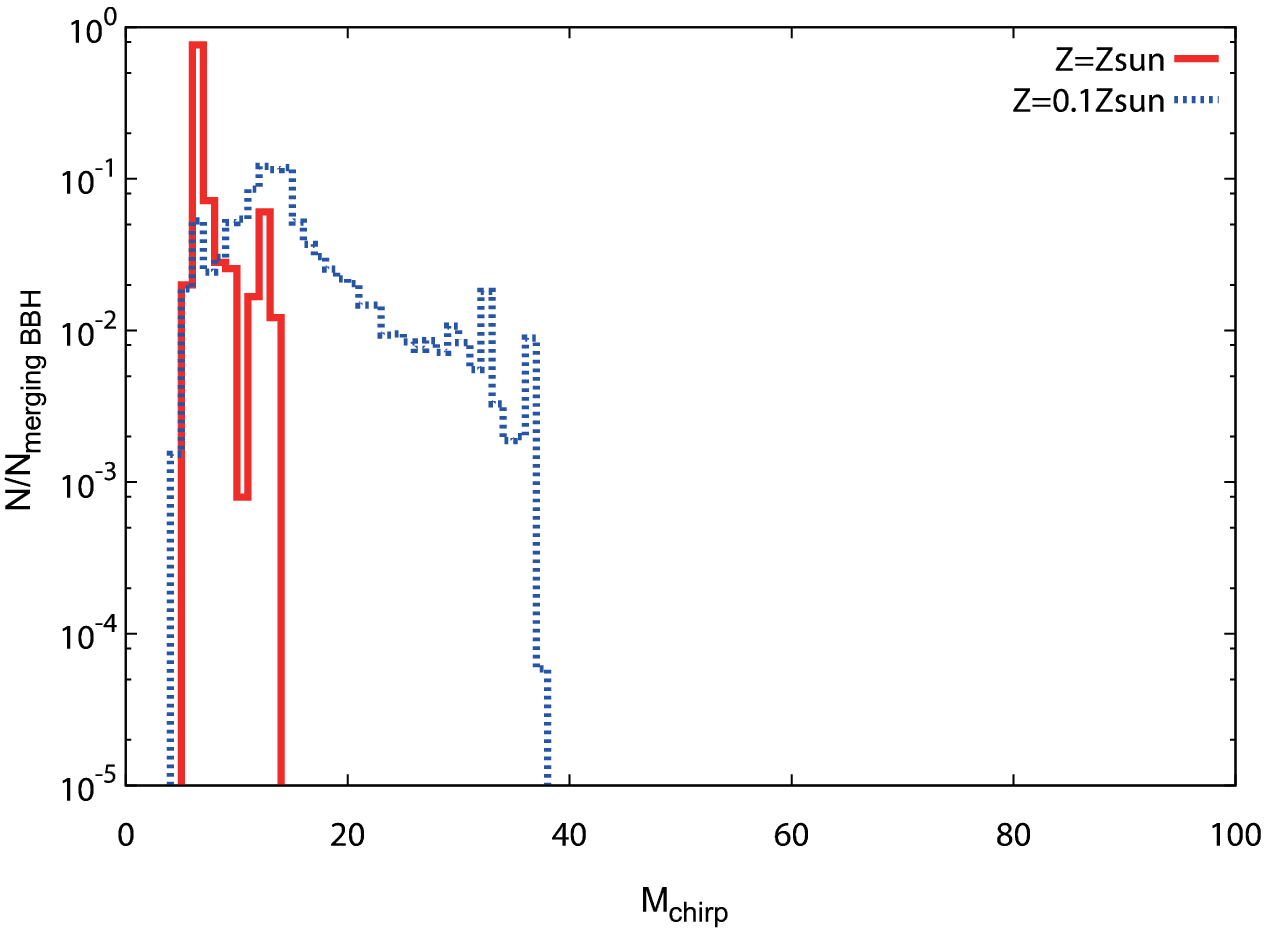}
\includegraphics[width=0.49\textwidth,clip=true]{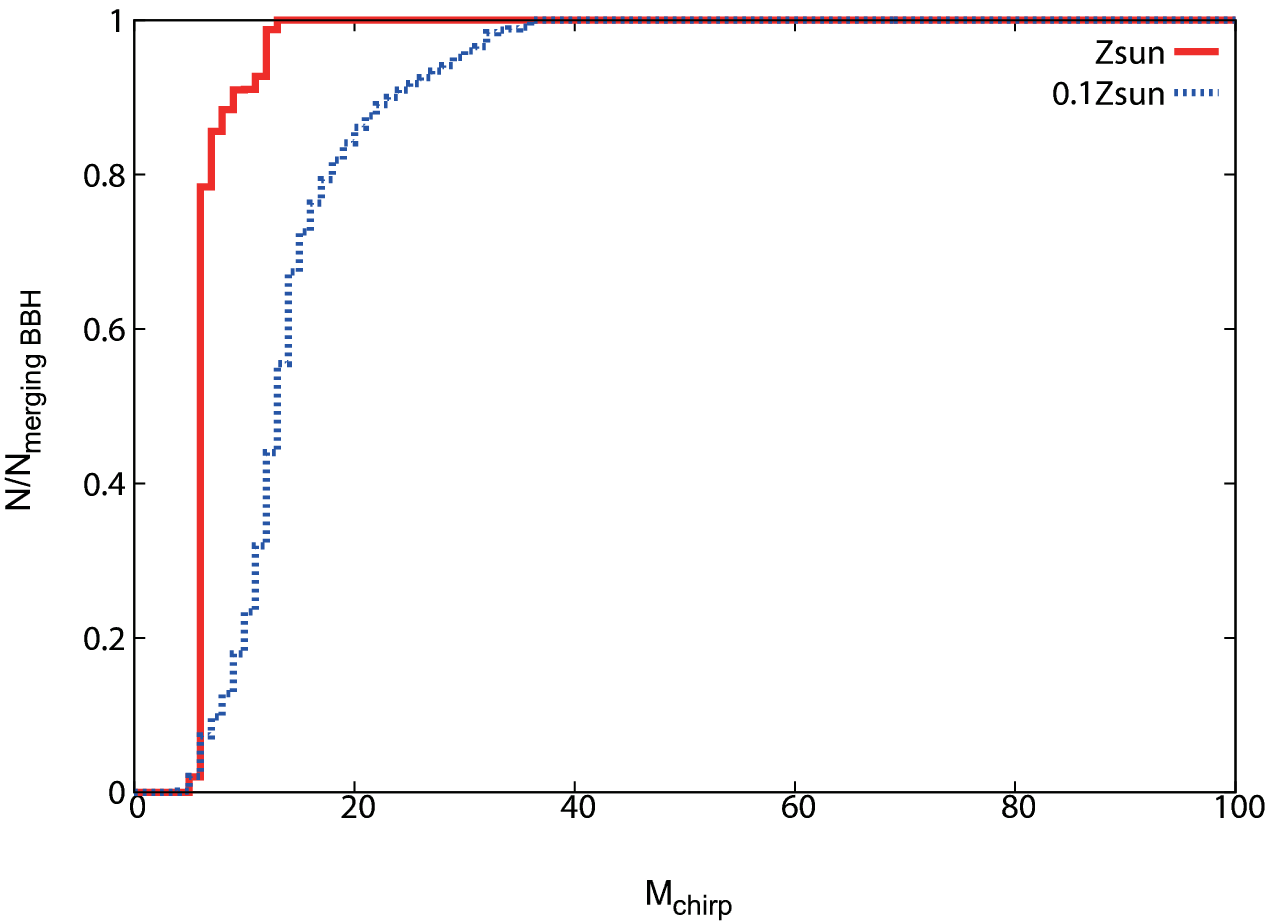}
\end{center}
 \caption{The chirp mass distributions of BBHs with the $Z=Z_\odot$ model
 and $Z=0.1Z_\odot$ model (left) and the cumulative chirp mass distribution of BBHs
 with the $Z=Z_\odot$ model and $Z=0.1Z_\odot$ model
 (right)~\cite{Dominik:2012kk}.}
\label{fig:PopIandIImass}
\end{figure}

It is considered that the stellar wind mass loss rate is an increasing function
of metallicity~\cite{Vink:2014qua},
i.e., Pop II binaries tend to be more massive compact binaries than Pop I.
Thus, the maximum mass of Pop II BBHs is greater than Pop I
BBHs.~\footnote{This result depends on the stellar wind mass loss
model~\cite{Belczynski:2009xy}.}
In the case of Pop II BBHs, however, the peak of the chirp mass distribution
is almost the same as that of Pop I, assuming that the Pop II IMF
is the same as the Pop I IMF. 
The reason is that in the late evolution stage,
Pop II stars evolve not as a blue giant like a Pop III star with mass less than
$50 M_\odot$ but as a red giant like a Pop I star.
Figure~\ref{fig:PopIandIImass} shows the chirp mass distribution
and the cumulative chirp mass distribution of Dominik's standard model of 
submodel B for $Z=Z_\odot$ and $Z=0.1Z_\odot$~\cite{Dominik:2012kk}~\footnote{
http://www.syntheticuniverse.org/ .}
(see also Fig. S6 in ``Supplementary online text''
of Belczynski's latest paper~\cite{Belczynski:2016obo}).

The top right panel of Fig.~\ref{fig:CER} shows the cumulative merger rates of
Belczynski's latest model~\cite{Belczynski:2016obo} and
previous model~\cite{Dominik:2013tma}.
In these models they use a grid of 11 metallicities such as
$Z=0.03,\,0.02\,(=Z_\odot),\,0.015,\,0.01,\,
0.005,\,0.002,\,0.0015,\,0.001,\,0.0005,\,0.0002,\,0.0001$. 
The cumulative merger rates of Pop I and II saturate at $z\leq6$.
Note that the cumulative merger rate depends on the SFR
and the metallicity evolution models.

\subsection{PBBH model}

Nakamura et al.~\cite{Nakamura:1997sm}~\footnote{
There are so many typos in this ApJ paper.
$x$ in Eq.~(1),  $x/x$ in Eq.~(3), $x^4/x^3$ in Eq.~(4),
$x  < y < x$ before Eq.~(7), $x$ in Eq.~(7), $y < x$ after Eq.~(7),
and $\frac{a}{x}$
in the definition of $e_{\rm max}$ and in Eq.~(8) should be $\bar{x}$, $x/\bar{x}$,
$x^4/\bar{x}^3$, $x  < y < \bar{x}$, $\bar{x}$, $y < \bar{x}$,
and $\frac{a}{\bar{x}}$, respectively. In the arXiv version,
astro-ph/9708060, there are no typos.}
first pointed out that the PBH can be
the source of GWs since many binary BHs can be formed by the existence
of the third BH.
In that paper, the prime interest was the possibility of PBH as
MACHO (MAssive Compact Halo Object) of mass
$\sim 0.5 M_\odot$~\cite{Yokoyama:1995ex,Jedamzik:1996mr}
as dark matter and the source of GW for TAMA300~\cite{Tagoshi:2000bz,Ando:2001ej}.
Sasaki et al.~\cite{Sasaki:2016jop} applied this to GW150914-like BBHs
(see also Refs.~\cite{Bird:2016dcv,Clesse:2016vqa}). 

If there exists random density perturbation in the radiation-dominated era,
in some part of the universe the amplitude of the density perturbation can be
high enough to collapse to a BH. Since the sound velocity is comparable to
the light velocity in the radiation-dominated era,
the Jeans length is comparable to the particle horizon size so that the mass of PBH
is comparable to the horizon mass at that time.
Then PBH with mass $\sim 30 M_\odot$ was formed at $\sim0.005$\,s and
the temperature of the universe was $\sim 14$\,MeV. Once PBH is formed,
it behaves like dust. Note here that the fraction of radiation
which collapses to PBH is only $\sim f\times 10^{-7.5}$ at the formation time
where $f$ is the fraction of PBH to the total mass of dark matter at present.
Ricotti, Ostriker, and Mack~\cite{Ricotti:2007au}
argued that if PBH exists there will be an accretion disk around PBH
and CMB will be distorted. For $\sim 30 M_\odot$ PBH, $f$ is at most $10^{-3}$
to be compatible with the data of COBE FIRAS~\cite{Fixsen:1996nj}.

After the equal time between the local BH energy density and
the radiation energy density, some pairs of PBHs with separation
$x$ smaller than $f^{1/3}\bar{x}$, where $\bar{x}$ is the mean separation of PBHs,
will merge in a free fall time during the globally radiation-dominated era.
Two PBHs acquire angular momentum by the tidal force of the third PBH
and they become a binary BH.
Following the argument of Nakamura et al.~\cite{Nakamura:1997sm},
Sasaki et al.~\cite{Sasaki:2016jop} obtained the distribution function of
the semi-major axis $a$ and the eccentricity $e$ given by
\begin{equation}
 dP(a,e)= \frac{3}{2}\, f^{3/2}\bar{x}^{-3/2}a^{1/2}e\,(1-e^2)^{-3/2} da\, de \,.
\label{eq:dP}
\end{equation}
The merger time $t$ due to the emission of the GW is approximately given
by Refs.~\cite{Peters:1963ux,Peters:1964zz}:
\begin{equation}
 t = C_a a^4(1-e^2)^{7/2} \,; \quad
 C_a = \frac {3}{170 c}\left(\frac{GM}{c^2}\right)^{-3} \,.
\end{equation}
We can integrate Eq.~\eqref{eq:dP} fixing the merger time $t$ to get
the probability of merger from $t$ to $t+dt$ as~\footnote{
When the BH fraction is even smaller, the upper bound of the 
integral over $e$ is determined differently. As a result, 
we obtain a slightly different expression for the merger 
probability~\cite{Sasaki:2016jop}.
}
\begin{equation}
 dP(t)=\frac{3}{29} \left( \left(\frac{t}{t_{\rm max}}\right)^{3/37}
 -\left(\frac{t}{t_{\rm max}}\right)^{3/8} \right)\frac{dt}{t} \,; \quad
 t_{\rm max}=\frac{\bar{x}^4C_a}{f^4} \,,
\end{equation}
while there is an upper bound of $e$ as
\begin{equation}
e_{\rm max}=\sqrt{1-\left(f \,\frac{a}{\bar{x}}\right)^{3/2}} \,.
\end{equation}

Now let us calculate the cumulative event rate ($R(z)$) as a function of $z$ as
\begin{eqnarray}
 \frac{dt}{dz}&=&-\frac{1}{H(z)(1+z)H_0} \,, \\
 \frac{dr}{dz}&=&\frac{c}{H(z)H_0} \,, \\
 \frac{dR}{dz}&=&4\pi \frac{r(z)^2}{H(z)H_0(1+z)}
 \left(\frac{\rho_c}{60M_\odot}\right)\Omega_{\rm DM}f\frac{dP}{dt} \,, \\
 H(z)&=&\sqrt{\Omega_{\rm DM}(1+z)^3+(1-\Omega_{\rm DM})}\\
 H_0&=&{\rm 70\,km~s^{-1}~Mpc^{-1}} \,,\\
 \Omega_{\rm DM}&=&0.3 \,,\\
 \rho_c&=&\frac{3H_0^2}{8\pi G} \,,
\end{eqnarray}
where we fixed the starting time $t_0=13.7$\,Gyr from the big bang.
The bottom panel of Fig.~\ref{fig:CER}
shows the cumulative event rate per year of PBBH mergers
as a function of $z$,
corresponding to the fraction $f=0.001$; that is, PBH contributes only 0.1\%
of the dark matter. Note that the rate is still increasing logarithmically
even at $z=30$.

\subsection{QNM, or ringing tail}

For the LISA detector,
a detailed study of ringdown GWs from massive BHs was done
by Berti, Cardoso, and Will~\cite{Berti:2005ys}.
Here, we model the ringdown waveform as
\bea
h(f_c,\,Q,\,t_0,\,\phi_0;\,t) \propto
\begin{cases}
e^{ - 2 \pi \,|f_I| \,(t-t_0)}\,\cos(2\,\pi \,f_R\,(t-t_0)-\phi_0) 
& {\rm for} \quad t \geq t_0 \,, \\
0 & {\rm for} \quad t < t_0 \,,
\end{cases}
\eea
where $f_R$ and $f_I$ denote the real and imaginary parts
of the QNM frequency, respectively,
and $t_0$ and $\phi_0$ are the initial ringdown time and phase, respectively.
For non-spinning, equal mass BBH mergers, $f_R$ and $f_I$ are given by
\bea
 f_R =299.5 \,{\rm Hz}\,
 (1+z)^{-1} \left(\frac{M_t}{60 M_\odot}\right)^{-1} \,, \quad 
 f_I = -46.34 \,{\rm Hz}\,(1+z)^{-1} \left(\frac{M_t}{60 M_\odot}\right)^{-1} \,,
\eea
where we used fitting formulas given in Ref.~\cite{Healy:2014yta}
(see also Refs.~\cite{Lousto:2009mf,Barausse:2012qz,Hofmann:2016yih})
to obtain the remnant BH parameters, i.e., the mass
and spin parameter of the Kerr BH,
and the BH parameters were converted to $f_R$ and $f_I$ by using Ref.~\cite{Berti:2005ys}.

As for the initial amplitude of the ringdown GWs,
we consider a simple inspiral--merger--ringdown waveform for BBHs
in the frequency domain, based on Ref.~\cite{Ajith:2009bn}.
In practice, thanks to breakthrough of numerical relativity
for BBHs~\cite{Pretorius:2005gq,Campanelli:2005dd,Baker:2005vv},
we can have precise inspiral--merger--ringdown waveforms for BBHs
(see, e.g., Refs~\cite{Ajith:2012az,Hinder:2013oqa,Abbott:2016apu}).
When we focus on the amplitudes for the three phases,
these are described by
\bea
 A(f) = C 
\begin{cases}
f^{-7/6} & {\rm for}~f < f_1 \,, \\
A_{\rm mer} f^{-2/3} & {\rm for}~ f_1 \leq f < f_2 \,, \\
\displaystyle{\frac{A_{\rm ring}}{(f-f_R)^2+f_I^2}} & {\rm for}~ f_2 \leq f \,,
\end{cases}
\eea
where we set $f_1=c^3/(6^{3/2}\pi G (1+z) M_t)$, which is 
twice the frequency of the innermost stable circular orbit (ISCO)
of a test particle in the Schwarzschild spacetime,
and $f_2=f_R$.
The constant $C$ is chosen to be consistent with the amplitude of the inspiral phase,
and $A_{\rm mer}$ and $A_{\rm ring}$ are calculated 
by imposing the continuity of $A(f)$ at $f=f_1$ and $f=f_2$, respectively.

\begin{figure}[!t]
\begin{center}
 \includegraphics[width=0.6\textwidth,clip=true]{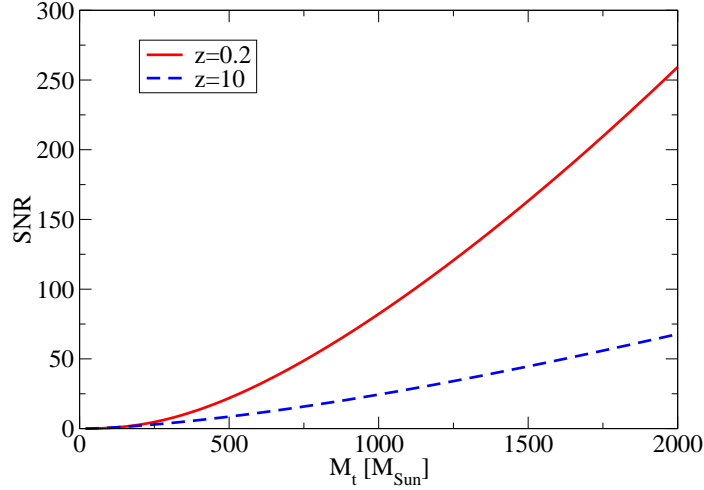}
\end{center}
 \caption{The total mass $M_t$ dependence of SNR for the ringdown phase
 where we assume the Pre-DECIGO noise curve.
 The solid (red) and dashed (blue) curves show the $z=0.2$ and $z=10$ cases, respectively.
 For simplicity, we have assumed non-spinning, equal mass BBHs.}
 \label{fig:QNM_pD}
\end{figure}

Using the amplitude $A_{\rm ring}$, we evaluate the SNR for the ringdown phase.
In Fig.~\ref{fig:QNM_pD}, we show the total mass $M_t$ dependence of
SNR in the Pre-DECIGO noise curve.
It is noted that for non-spinning, equal mass binaries,
the ringdown event with SNR~$=35$ is enough to confirm Einstein's GR
for the strong gravity space-time near the event horizon,
i.e., whether the compact object emitting the ringdown GWs
is a BH predicted by GR or not (see Refs.~\cite{Berti:2007zu,Nakano:2015uja}).
As shown in Fig.~\ref{fig:QNM_pD}, we have the possibility
of testing general relativity to that accuracy
for $M_t > 640M_{\odot}$ and $1280M_{\odot}$ at $z=0.2$ and $10$, respectively.

\subsection{Accuracy of direction and time of the merger of BBH}

In this section, we investigate how accurately
we can determine the waveform parameters 
of GWs from GW150914-like BBHs using the Fisher analysis. 
Parameter resolutions are estimated by the inverse of Fisher matrix
$\Gamma_{ij}$ for large SNR as~\cite{Finn:1992wt, Cutler:1994ys}
\begin{align}
 \langle \Delta \theta^i \Delta \theta^j \rangle = \left(\Gamma^{-1} \right)_{ij} \,, 
\label{Eq:ParamsResol}
\end{align}
where $\Gamma_{ij}$ is defined as 
\begin{align}
 \Gamma_{ij} = \left( \dfrac{\partial h}{\partial \theta^i}
 \Big| \dfrac{\partial h}{\partial \theta^j} \right) \,.
\end{align}
The symbol $\left(\cdot | \cdot\right)$ denotes the noise-weighted inner product. 
Assuming stationary Gaussian detector noise, 
the inner product is expressed by 
\begin{align}
 \left(A|B\right) \equiv 4 \text{Re} \int_{f_{\text{ini}}}^{f_{\text{fin}}} df
 \dfrac{\tilde{A}\left(f\right) \tilde{B}^{\ast} \left(f\right)}{S_n\left(f\right)}
 \,.
\label{Eq:innerProduct}
\end{align}
We define the accuracy of sky localization as the measurement error
in the solid angle, 
\begin{align}
 \Delta \Omega 
 \equiv 2\pi \left| \sin \delta \right|
 \sqrt{ \langle\Delta\alpha^2\rangle \langle\Delta \delta^2\rangle
 - \langle\Delta\alpha\Delta\delta\rangle^2 } \,,
\label{Eq:AngularResolution}
\end{align}
where $\alpha$ and $\delta$ denote the right ascension
and declination of the GW source, respectively. 

We consider a GW from a coalescing binary system consisting of two point non-spinning masses with $M_1$ and $M_2$. 
We use the restricted post-Newtonian (PN) waveform up to the
1.5PN order~\cite{Cutler:1997ta}: 
\begin{align}
 &\tilde{h} \left(f\right)
 =\mathcal{A} Q\left(t \left(f\right) \right)f^{-7/6}
 e^{-i\left[\phi_{p}\left(t\left(f\right)\right)
 + \phi_{D}\left(t\left(f\right)\right)+\Psi\left(f\right)\right]} \,, 
\label{Eq:hGW} \\
 &\mathcal{A} = \sqrt{\dfrac{5}{24}} \dfrac{1}{\pi^{2/3}} \dfrac{c}{d_L(z)}
 \left(  \dfrac{\left(1+z\right)GM_c}{c^3} \right)^{5/6} \,, 
\\
 &Q\left(t \left(f\right) \right) = 
 \left[ \left(\dfrac{1 + \cos^2\iota}{2} \right)^2 
 F_+\left(t \left(f\right),\alpha,\delta\right)^2 
 + \cos^2\iota F_{\times}\left(t \left(f\right),\alpha,\delta\right)^2 \right]^{1/2} 
 \,,
\\
 &\phi_p \left(t\left(f\right)\right) 
 = \arctan \left[ -\dfrac{2\cos\iota}{1+\cos^2\iota}
 \dfrac{F_{\times}\left(t \left(f\right),\alpha,\delta\right)}
 {F_+\left(t \left(f\right),\alpha,\delta\right)} \right] \,,
\\
 &\phi_D\left(t\left(f\right)\right)
 = 2\pi f  \dfrac{\boldsymbol{n}\cdot \boldsymbol{r}
 \left( t \left(f\right)\right)}{c} \,,
\\
 &\Psi\left(f\right)
 = 2\pi f t_c - \phi_c - \dfrac{\pi}{4}
 + \dfrac{3}{4} \left( \dfrac{8\pi G\left(1+z\right)M_c}{c^3} \right)^{-5/3}
 \left[ 1 + \dfrac{20}{9} \left( \dfrac{743}{336}  +\dfrac{11}{4}\eta \right)x
 - 16\pi x^{3/2} \right] \,,
\\
 &t\left(f\right)
 = t_c - 5\left(\dfrac{G\left(1+z\right)M_c}{c^3}\right)^{-5/3}
 \left(8\pi f\right)^{8/3}
 \left[ 1 + \dfrac{4}{3} \left( \dfrac{743}{336}  +\dfrac{11}{4}\eta \right)x
 - \dfrac{32}{5} \pi x^{3/2} \right] \,,
\label{Eq:timeGW}
\end{align} 
where the PN parameter
$x \equiv \left[ \pi G \left(1+z\right) \left(M_1 + M_2\right)f/c^3 \right]^{2/3}$
was introduced. 
$F_+$ and $F_\times$ denote the ``detector beam-pattern'' coefficients,
$\iota$ and $\phi_c$ are the inclination angle of the binary
and the phase at the coalescence time $t_c$, respectively,
and $\eta=M_1M_2/(M_1+M_2)^2$.
The vectors $\boldsymbol{n}$ and $\boldsymbol{r}$ appearing
in the Doppler phase $\phi_{\text{D}}$ are 
the unit vector pointing from the geometrical center of Pre-DECIGO 
to the source and the vector pointing from the SSB (Solar System barycenter)
to the geometrical center of Pre-DECIGO, respectively. 

\begin{figure}[!t]
\begin{center}
\includegraphics[width=1.0\textwidth,clip=true]{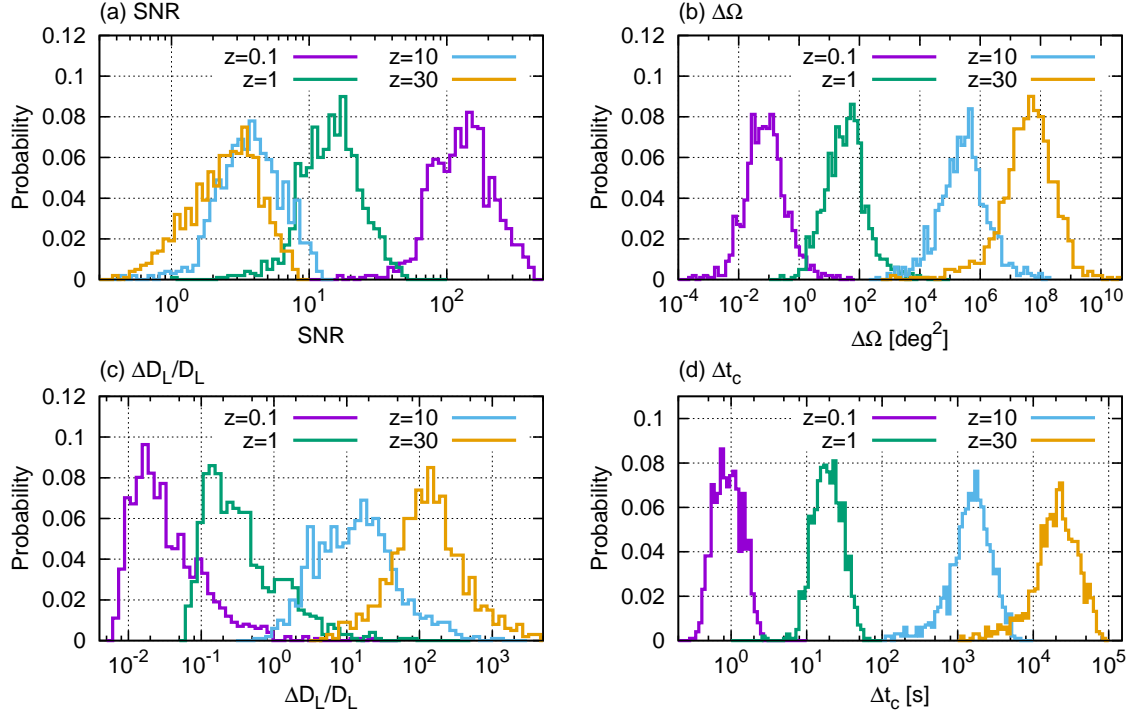}
\end{center}
 \caption{Histograms for (a) SNR, (b) angular resolution, 
 (c) luminosity distance, and (d) coalescence time 
 of GWs from $30 M_{\odot}$ equal mass BH binaries 
 obtained by using $10^3$ Monte Carlo simulations. 
 We assume that the sky location and the inclination angle are 
 distributed according to uniform distributions. 
 The polarization phase is set to be 0.5 radians. 
 The four lines correspond to the different redshifts $z=0.1,\, 1,\, 10$ and $30$.}
\label{Fig:ParamsResol}
\end{figure}

We investigate the angular resolution defined by Eq.~(\ref{Eq:AngularResolution})
and the accuracy of the time of coalescence using Fisher analysis. 
The GW signal is characterized by the nine waveform parameters 
$\boldsymbol{\theta} = \left\{\alpha, \delta, t_c, \phi_c, d_L, \ln M_c,
\ln \eta, \psi, \cos\iota \right\}$ (see Ref.~\cite{Cutler:1997ta} for details;
the polarization angle $\psi$ is in $F_+$ and $F_\times$).
We set the cutoff frequencies in Eq.~(\ref{Eq:innerProduct}) in the following way: 
\begin{align}
 &f_{\text{ini}} = \text{max} \left\{ f_{1\text{yr}}, 0.01 \,\text{Hz} \right\} \,, \\
 &f_{\text{fin}} = \text{min} \left\{f_{\text{ISCO}}, 10 \,\text{Hz} \right\} \,,
\end{align} 
where $f_{\text{ISCO}} \equiv c^3 / \left[6^{3/2} \pi G \left(1+z\right)
\left(M_1 + M_2\right) \right]$ is 
the GW frequency at the ISCO, 
and $f_{1\text{yr}}$ is the frequency at one year before the coalescence. 
Substituting Eqs.~(\ref{Eq:hGW})--(\ref{Eq:timeGW})
into Eq.~(\ref{Eq:ParamsResol}), we evaluate the accuracy of the parameter estimation. 
When we perform the Fisher analysis, we assume GW signals from $30 M_{\odot}$
equal-mass BBHs at distances of $z=0.1, \,1$ and $10$.
The polarization phase $\psi$ is set to be 0.5 radians. The sky location $\left(\alpha,\delta \right)$ 
and the inclination $\cos\iota$ are assumed to obey uniform distributions and are averaged over by using $10^3$ Monte Carlo simulations. 
As a result, we obtain Fig.~\ref{Fig:ParamsResol}. 

\begin{figure}[!b]  
\begin{center}
\includegraphics[width=1.0\textwidth,clip=true]{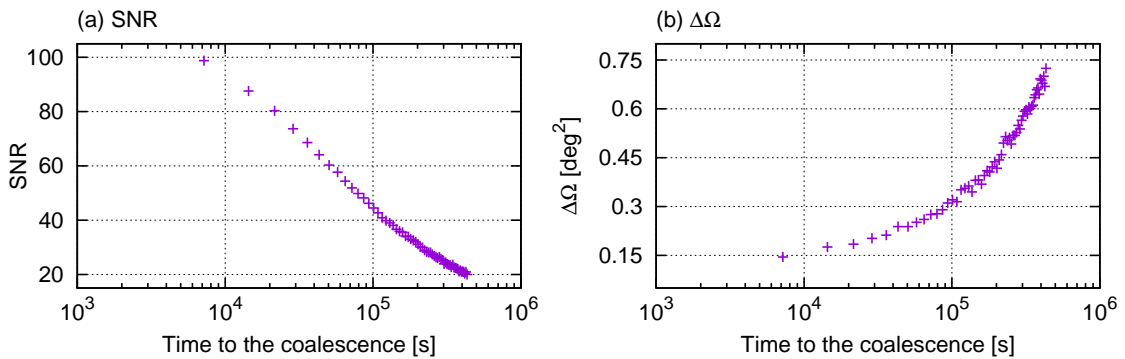}
\end{center}
 \caption{Time evolution for (a) SNR and (b) angular resolution
 of GWs from $30M_{\odot}$ equal mass BH binaries at the distance of $z=0.1$. 
 We assume $\alpha=\delta = 1.0$\,rad,
 $\psi=0.5$\,rad, and $\cos\iota=0.5$.}
\label{Fig:timeEvol_SNR_dOmega}
\end{figure}  

Figure~\ref{Fig:ParamsResol} shows the probability distributions for 
(a) SNR, (b) angular resolution, (c) luminosity distance, and (d) coalescence time. 
As can be seen in the top left panel of this figure, GW150914-like GW signals 
are typically detectable by Pre-DECIGO with SNR of 130 and 15
for $z=0.1$ and 1, respectively. 
Even for GWs from $z=30 \ (10)$, 
about 10\% of them can be observed by Pre-DECIGO with $\text{SNR} >5 \ (7)$,
thanks to the fact that the overall amplitude only depends on
the redshift as $z^{-1/6}$ for $z \gg 1$. 
Figure~\ref{Fig:ParamsResol} (b) shows that we can typically localize
the sky position of the source to 
$\Delta \Omega = 7.6 \times 10^{-2}\,{\rm deg}^2$
and $35\,{\rm deg}^2$ for $z=0.1$ and $1$, respectively. 
These values are easily estimated by Eq.~(51) in Ref.~\cite{Wen:2010cr} as 
\begin{align}
 \Delta \Omega \simeq 35 \, {\rm deg}^2 
 \left( \dfrac{0.1 \, \text{Hz}}{f}\right)^2 \left( \dfrac{10}{\text{SNR}} \right)^2 
 \left( \dfrac{10^6 \,\text{s}}{T} \right)^3 \,,
\end{align}
for $z=1$ where the detector trajectory is assumed
to be a circular orbit around the Sun. 
Figure~\ref{Fig:timeEvol_SNR_dOmega} shows the time evolution of the SNR
and the angular resolution of GWs from $z=0.1$.
The angle parameters are set to be 
$\alpha=\delta = 1.0$\,rad,
$\psi=0.5$\,rad, and $\cos\iota=0.5$.  
This figure indicates that we can identify the sky location of GW150914-like BBHs 
with an accuracy of about $0.3\,{\rm deg}^2$
at about a day before the coalescence. 
Note that since SNRs are not so large in the case of $z=10$ and $30$, 
the estimation accuracies of the parameters would be
overestimated \cite{Balasubramanian:1995bm}.

\section{Discussion}

One of the observed binary NS PSR2127+11C 
(see Ref.~\cite{Jacoby:2006dy} and references therein)
has a similar parameter to
the Hulse--Taylor binary pulsar although PSR2127+11C
is in the globular cluster (GC) M15.
This suggests the possibility of the formation of BBHs in the GC.
A BH of mass $\sim 30 M_\odot$ is much larger than the typical mass of 
the constituent stars, $\sim 1 M_\odot$,
so that it will sink down to the center of the GC or star cluster 
due to dynamical friction. Then BBHs can be formed
in the central high density region of GCs. Since the escape velocity from GCs is
$10\,{\rm km\, s^{-1}}$ or so, the kick velocity in the formation process of BHs
or the kick when BBHs are formed by three-body interaction is
high enough for BBHs to escape from GCs.
Rodriguez, Chatterjee, and Rasio~\cite{Rodriguez:2016kxx}
performed such a simulation to show that the event rate is at most $\sim 1/7$
of Pop I and II origin BBHs. 
If we take their result as it is,
the dynamical formation of binaries in GCs gives only
a minor contribution of Pop II origin of BBHs.

\begin{figure}[!t]
\begin{center}
 \includegraphics[width=0.6\textwidth,clip=true]{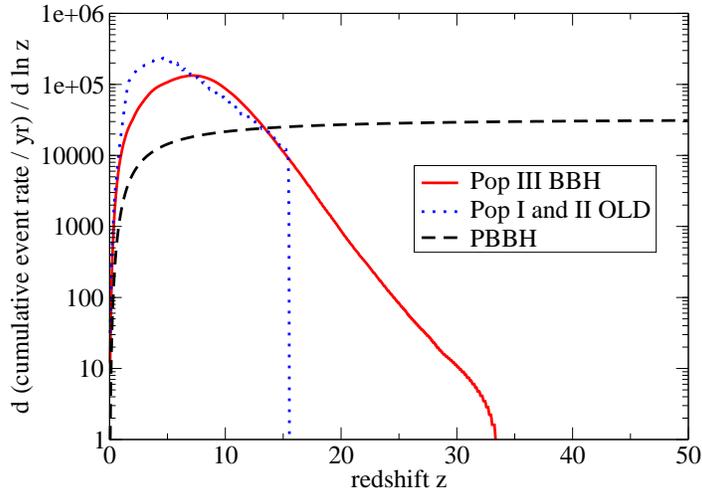}
\end{center}
\caption{The event rates for Pop III (standard),
Pop I and II (OLD), and PBBH merger as a function of $z$.
These rates are derived by differentiating
the cumulative event rate in Fig.~\ref{fig:CER} with respect to $\ln z$.
Note here that the detectability may change
by the mass distribution of each model.} 
\label{fig:combine3}
\end{figure}
  
From only the chirp mass, total mass and spin angular momentum,
it will be difficult to distinguish the origin of GW150914-like BBHs.
This is because the number of parameters that can be determined
by the distribution function of the GW data
is much smaller than that of the unknown model parameters and
the distribution functions assumed in each model.
However, the redshift distribution of GW events varies
robustly among the models.
Namely, the maximum possible redshift is $\sim 6,\,10$, and $> 30$
for Pop I/II, Pop III, and PBBH models, respectively (see Fig.~\ref{fig:combine3}).
In Fig.~\ref{fig:combine3}, we show the event rates for each model.
These event rates are derived by differentiating
the cumulative event rate in Fig.~\ref{fig:CER} with respect to $\ln z$.
To observe the maximum redshift as a smoking gun to identify the origin
of GW150914-like events,
the construction of Pre-DECIGO seems to be the unique possibility.

Pre-DECIGO can observe NS--NS and NS--BH mergers.
However no detection of GWs from the merger of these systems
has been done, though many simulations exist.
For the same distance of the source, the SNR for NS--NS and NS--BH (30$M_\odot$)
are 0.08 and 0.25 times smaller than for $30M_\odot$--$30M_\odot$ BBHs.
We will here postpone discussing what we can do using Pre-DECIGO
about these sources until the first observations of GWs from
these systems, since the event rates are still uncertain and might be
very small compared with BBH mergers.

\section*{Acknowledgments}

The authors would like to thank A.~Miyamoto and T.~Suyama for
careful reading of the manuscript.
This work was supported by the Grant-in-Aid from the Ministry of Education,
Culture, Sports, Science and Technology (MEXT) of Japan No.~15H02087
and by MEXT Grant-in-Aid for Scientific Research on Innovative Areas,
``New Developments in Astrophysics Through Multi-Messenger Observations
of Gravitational Wave Sources,'' No.~24103006 (TN, TT, NS, HN), 
JSPS Grant-in-Aid for Scientific Research (C), No.~16K05347 (HN), 
JSPS Fellows Grant No.~26.8636 (KE), and 
JSPS Grants-in-Aid for Scientific Research (KAKENHI) 15H02082, 24103005,
and 15K05070 (YI).



\begin{thebibliography}{99}

\bibitem{Abbott:2016blz} 
  B.~P.~Abbott {\it et al.} [LIGO Scientific and Virgo Collaborations],
  Phys.\ Rev.\ Lett.\  {\bf 116}, 061102 (2016)
  [arXiv:1602.03837 [gr-qc]].

\bibitem{Kalogera:2006uj} 
  V.~Kalogera, K.~Belczynski, C.~Kim, R.~W.~O'Shaughnessy and B.~Willems,
  Phys.\ Rept.\  {\bf 442}, 75 (2007)
  [astro-ph/0612144].

\bibitem{Dominik:2012kk} 
  M.~Dominik, K.~Belczynski, C.~Fryer, D.~Holz, E.~Berti, T.~Bulik, I.~Mandel and R.~O'Shaughnessy,
  Astrophys.\ J.\  {\bf 759}, 52 (2012)
  [arXiv:1202.4901 [astro-ph.HE]].

\bibitem{Kinugawa:2014zha} 
  T.~Kinugawa, K.~Inayoshi, K.~Hotokezaka, D.~Nakauchi and T.~Nakamura,
  Mon.\ Not.\ Roy.\ Astron.\ Soc.\  {\bf 442}, 2963 (2014)
  [arXiv:1402.6672 [astro-ph.HE]].

\bibitem{Spera:2015vkd} 
  M.~Spera, M.~Mapelli and A.~Bressan,
  Mon.\ Not.\ Roy.\ Astron.\ Soc.\  {\bf 451}, 4086 (2015)
  [arXiv:1505.05201 [astro-ph.SR]].

\bibitem{Kinugawa:2015nla} 
  T.~Kinugawa, A.~Miyamoto, N.~Kanda and T.~Nakamura,
   Mon.\ Not.\ Roy.\ Astron.\ Soc.\  {\bf 456}, 1093 (2016)
   [arXiv:1505.06962 [astro-ph.SR]].

\bibitem{Amaro-Seoane:2015umi} 
  P.~Amaro-Seoane and X.~Chen,
  Mon.\ Not.\ Roy.\ Astron.\ Soc.\  {\bf 458}, 3075 (2016)
  [arXiv:1512.04897 [astro-ph.CO]].

\bibitem{Mandel:2015qlu} 
  I.~Mandel and S.~E.~de Mink,
  Mon.\ Not.\ Roy.\ Astron.\ Soc.\  {\bf 458}, 2634 (2016)
  [arXiv:1601.00007 [astro-ph.HE]].

\bibitem{Marchant:2016wow} 
  P.~Marchant, N.~Langer, P.~Podsiadlowski, T.~M.~Tauris and T.~J.~Moriya,
  Astron.\ Astrophys.\  {\bf 588}, A50 (2016)
  [arXiv:1601.03718 [astro-ph.SR]].

\bibitem{Belczynski:2016obo} 
  K.~Belczynski, D.~E.~Holz, T.~Bulik and R.~O'Shaughnessy,
  arXiv:1602.04531 [astro-ph.HE].

\bibitem{Kim:2002uw} 
  C.~Kim, V.~Kalogera and D.~R.~Lorimer,
  Astrophys.\ J.\  {\bf 584}, 985 (2003)
  [astro-ph/0207408].

\bibitem{Kalogera:2003tn} 
  V.~Kalogera {\it et al.},
  Astrophys.\ J.\  {\bf 601}, L179 (2004)
  [Astrophys.\ J.\  {\bf 614}, L137 (2004)]
  [astro-ph/0312101].

\bibitem{Kim:2013tca} 
  C.~Kim, B.~B.~P.~Perera and M.~A.~McLaughlin,
  Mon.\ Not.\ Roy.\ Astron.\ Soc.\  {\bf 448}, 928 (2015)
  [arXiv:1308.4676 [astro-ph.SR]].

\bibitem{Abadie:2010cf} 
  J.~Abadie {\it et al.} [LIGO Scientific and VIRGO Collaborations],
  Class.\ Quant.\ Grav.\  {\bf 27}, 173001 (2010)
  [arXiv:1003.2480 [astro-ph.HE]].

\bibitem{Baraffe:2000dp} 
  I.~Baraffe, A.~Heger and S.~E.~Woosley,
  Astrophys.\ J.\  {\bf 550}, 890 (2001)
  [astro-ph/0009410].

\bibitem{Inayoshi:2013rfa} 
  K.~Inayoshi, T.~Hosokawa and K.~Omukai,
  Mon.\ Not.\ Roy.\ Astron.\ Soc.\  {\bf 431}, 3036 (2013)
  [arXiv:1302.6065 [astro-ph.SR]].

\bibitem{Kinugawa:2016mfs} 
  T.~Kinugawa, H.~Nakano and T.~Nakamura,
  Prog. Theor. Exp. Phys. (2016), 031E01
  [arXiv:1601.07217 [astro-ph.HE]].

\bibitem{Kinugawa:2016ect} 
  T.~Kinugawa, H.~Nakano and T.~Nakamura,
  arXiv:1606.00362 [astro-ph.HE].

\bibitem{TheLIGOScientific:2016htt} 
  B.~P.~Abbott {\it et al.} [LIGO Scientific and Virgo Collaborations],
  Astrophys.\ J.\  {\bf 818}, L22 (2016)
  [arXiv:1602.03846 [astro-ph.HE]].

\bibitem{Sasaki:2016jop} 
  M.~Sasaki, T.~Suyama, T.~Tanaka and S.~Yokoyama,
  Phys.\ Rev.\ Lett.\  {\bf 117}, 061101 (2016)
  [arXiv:1603.08338 [astro-ph.CO]].

\bibitem{Nakamura:1997sm} 
  T.~Nakamura, M.~Sasaki, T.~Tanaka and K.~S.~Thorne,
  Astrophys.\ J.\  {\bf 487}, L139 (1997)
  [astro-ph/9708060].

\bibitem{Seto:2001qf} 
  N.~Seto, S.~Kawamura and T.~Nakamura,
  Phys.\ Rev.\ Lett.\  {\bf 87}, 221103 (2001)
  [astro-ph/0108011].

\bibitem{Kawamura:2006up} 
  S.~Kawamura {\it et al.},
  Class.\ Quant.\ Grav.\  {\bf 23}, S125 (2006).

\bibitem{Punturo:2010zz} 
  M.~Punturo {\it et al.},
  Class.\ Quant.\ Grav.\  {\bf 27}, 194002 (2010).

\bibitem{KAGRA:1999} 
  K.~Kuroda {\it et al.} [LCGT Collaboration],
  Int.\ J.\ Modern\ Phys.\  {\bf D8}, 557 (1999).

\bibitem{ET:2011} 
 {\it Einstein gravitational wave Telescope Conceptual Design Study},
 http://www.et-gw.eu/
 (2011).

\bibitem{eLISA:2016} 
 {\it The ESA-L3 GravitationalWaveMission
 Gravitational Observatory Advisory Team Final Report},
 http://www.cosmos.esa.int/web/goat  
 (2016).

\bibitem{Dalal:2006qt} 
  N.~Dalal, D.~E.~Holz, S.~A.~Hughes and B.~Jain,
  Phys.\ Rev.\ D {\bf 74}, 063006 (2006)
  [astro-ph/0601275].

\bibitem{Vallisneri:2012np} 
  M.~Vallisneri and C.~R.~Galley,
  Class.\ Quant.\ Grav.\  {\bf 29}, 124015 (2012)
  [arXiv:1201.3684 [gr-qc]].

\bibitem{deSouza:2011ea} 
  R.~S.~de Souza, N.~Yoshida and K.~Ioka,
  Astron.\ Astrophys.\  {\bf 533}, A32 (2011)
  [arXiv:1105.2395 [astro-ph.CO]].

\bibitem{Omukai:2005hv} 
  K.~Omukai, T.~Tsuribe, R.~Schneider and A.~Ferrara,
  Astrophys.\ J.\  {\bf 626}, 627 (2005)
  [astro-ph/0503010].

\bibitem{Salpeter:1955it} 
  E.~E.~Salpeter,
  Astrophys.\ J.\  {\bf 121}, 161 (1955).

\bibitem{Kroupa:2000iv} 
  P.~Kroupa,
  Mon.\ Not.\ Roy.\ Astron.\ Soc.\  {\bf 322}, 231 (2001)
  [astro-ph/0009005].

\bibitem{Vink:2014qua} 
  J.~S.~Vink,
  in Vink J. S., ed., {\it Astrophysics and Space Science Library} 422, 
  {\it Very Massive Stars in the Local Universe}, 77
  arXiv:1406.5357 [astro-ph.SR].

\bibitem{Belczynski:2009xy} 
  K.~Belczynski, T.~Bulik, C.~L.~Fryer, A.~Ruiter, J.~S.~Vink and J.~R.~Hurley,
  Astrophys.\ J.\  {\bf 714}, 1217 (2010)
  [arXiv:0904.2784 [astro-ph.SR]].

\bibitem{Dominik:2013tma} 
  M.~Dominik, K.~Belczynski, C.~Fryer, D.~E.~Holz, E.~Berti, T.~Bulik, I.~Mandel and R.~O'Shaughnessy,
  Astrophys.\ J.\  {\bf 779}, 72 (2013)
  [arXiv:1308.1546 [astro-ph.HE]].

\bibitem{Yokoyama:1995ex} 
  J.~Yokoyama,
  Astron.\ Astrophys.\  {\bf 318}, 673 (1997)
  [astro-ph/9509027].

\bibitem{Jedamzik:1996mr} 
  K.~Jedamzik,
  Phys.\ Rev.\ D {\bf 55}, 5871 (1997)
  [astro-ph/9605152].

\bibitem{Tagoshi:2000bz} 
  H.~Tagoshi {\it et al.} [TAMA Collaboration],
  Phys.\ Rev.\ D {\bf 63}, 062001 (2001)
  [gr-qc/0012010].

\bibitem{Ando:2001ej} 
  M.~Ando {\it et al.} [TAMA Collaboration],
  Phys.\ Rev.\ Lett.\  {\bf 86}, 3950 (2001)
  [astro-ph/0105473].

\bibitem{Bird:2016dcv} 
  S.~Bird, I.~Cholis, J.~B.~Munoz, Y.~Ali-Haimoud, M.~Kamionkowski, E.~D.~Kovetz, A.~Raccanelli and A.~G.~Riess,
  Phys.\ Rev.\ Lett.\  {\bf 116}, 201301 (2016)
  [arXiv:1603.00464 [astro-ph.CO]].

\bibitem{Clesse:2016vqa} 
  S.~Clesse and J.~Garcia-Bellido,
  arXiv:1603.05234 [astro-ph.CO].

\bibitem{Ricotti:2007au} 
  M.~Ricotti, J.~P.~Ostriker and K.~J.~Mack,
  Astrophys.\ J.\  {\bf 680}, 829 (2008)
  [arXiv:0709.0524 [astro-ph]].

\bibitem{Fixsen:1996nj} 
  D.~J.~Fixsen, E.~S.~Cheng, J.~M.~Gales, J.~C.~Mather, R.~A.~Shafer and E.~L.~Wright,
  Astrophys.\ J.\  {\bf 473}, 576 (1996)
  [astro-ph/9605054].

\bibitem{Peters:1963ux} 
  P.~C.~Peters and J.~Mathews,
  Phys.\ Rev.\  {\bf 131}, 435 (1963).

\bibitem{Peters:1964zz} 
  P.~C.~Peters,
  Phys.\ Rev.\  {\bf 136}, B1224 (1964).

\bibitem{Berti:2005ys} 
  E.~Berti, V.~Cardoso and C.~M.~Will,
  Phys.\ Rev.\ D {\bf 73}, 064030 (2006)
  [gr-qc/0512160].

\bibitem{Healy:2014yta} 
  J.~Healy, C.~O.~Lousto and Y.~Zlochower,
  Phys.\ Rev.\ D {\bf 90}, 104004 (2014)
  [arXiv:1406.7295 [gr-qc]].

\bibitem{Lousto:2009mf} 
  C.~O.~Lousto, M.~Campanelli, Y.~Zlochower and H.~Nakano,
  Class.\ Quant.\ Grav.\  {\bf 27}, 114006 (2010)
  [arXiv:0904.3541 [gr-qc]].

\bibitem{Barausse:2012qz} 
  E.~Barausse, V.~Morozova and L.~Rezzolla,
  Astrophys.\ J.\  {\bf 758}, 63 (2012)
  [Astrophys.\ J.\  {\bf 786}, 76 (2014)]
  [arXiv:1206.3803 [gr-qc]].

\bibitem{Hofmann:2016yih} 
  F.~Hofmann, E.~Barausse and L.~Rezzolla,
  Astrophys.\ J.\  {\bf 825}, L19 (2016)
  [arXiv:1605.01938 [gr-qc]].

\bibitem{Ajith:2009bn} 
  P.~Ajith {\it et al.},
  Phys.\ Rev.\ Lett.\  {\bf 106}, 241101 (2011)
  [arXiv:0909.2867 [gr-qc]].

\bibitem{Pretorius:2005gq} 
  F.~Pretorius,
  Phys.\ Rev.\ Lett.\  {\bf 95}, 121101 (2005)
  [gr-qc/0507014].

\bibitem{Campanelli:2005dd} 
  M.~Campanelli, C.~O.~Lousto, P.~Marronetti and Y.~Zlochower,
  Phys.\ Rev.\ Lett.\  {\bf 96}, 111101 (2006)
  [gr-qc/0511048].

\bibitem{Baker:2005vv} 
  J.~G.~Baker, J.~Centrella, D.~I.~Choi, M.~Koppitz and J.~van Meter,
  Phys.\ Rev.\ Lett.\  {\bf 96}, 111102 (2006)
  [gr-qc/0511103].

\bibitem{Ajith:2012az} 
  P.~Ajith {\it et al.},
  Class.\ Quant.\ Grav.\  {\bf 29}, 124001 (2012)
  Addendum: [Class.\ Quant.\ Grav.\  {\bf 30}, 199401 (2013)]
  [arXiv:1201.5319 [gr-qc]].

\bibitem{Hinder:2013oqa} 
  I.~Hinder {\it et al.},
  Class.\ Quant.\ Grav.\  {\bf 31}, 025012 (2014)
  [arXiv:1307.5307 [gr-qc]].

\bibitem{Abbott:2016apu} 
  B.~P.~Abbott {\it et al.} [LIGO Scientific and Virgo Collaborations],
  arXiv:1606.01262 [gr-qc].

\bibitem{Berti:2007zu} 
  E.~Berti, J.~Cardoso, V.~Cardoso and M.~Cavaglia,
  Phys.\ Rev.\ D {\bf 76}, 104044 (2007)
  [arXiv:0707.1202 [gr-qc]].

\bibitem{Nakano:2015uja} 
  H.~Nakano, T.~Tanaka and T.~Nakamura,
  Phys.\ Rev.\ D {\bf 92}, 064003 (2015)
  [arXiv:1506.00560 [astro-ph.HE]].

\bibitem{Finn:1992wt} 
  L.~S.~Finn,
  Phys.\ Rev.\ D {\bf 46}, 5236 (1992)
  [gr-qc/9209010].
 
\bibitem{Cutler:1994ys} 
  C.~Cutler and E.~E.~Flanagan,
  Phys.\ Rev.\ D {\bf 49}, 2658 (1994)
  [gr-qc/9402014].
  
\bibitem{Cutler:1997ta} 
  C.~Cutler,
  Phys.\ Rev.\ D {\bf 57}, 7089 (1998)
  [gr-qc/9703068].

\bibitem{Wen:2010cr} 
  L.~Wen and Y.~Chen,
  Phys.\ Rev.\ D {\bf 81}, 082001 (2010)
  [arXiv:1003.2504 [astro-ph.CO]].

\bibitem{Balasubramanian:1995bm}
  R.~Balasubramanian, B.~S.~Sathyaprakash and S.~V.~Dhurandhar,
  Phys.\ Rev.\ D {\bf 53} (1996) 3033
   Erratum: [Phys.\ Rev.\ D {\bf 54} (1996) 1860]
  [gr-qc/9508011].

\bibitem{Jacoby:2006dy} 
  B.~A.~Jacoby, P.~B.~Cameron, F.~A.~Jenet, S.~B.~Anderson, R.~N.~Murty and S.~R.~Kulkarni,
  Astrophys.\ J.\  {\bf 644}, L113 (2006)
  [astro-ph/0605375].

\bibitem{Rodriguez:2016kxx} 
  C.~L.~Rodriguez, S.~Chatterjee and F.~A.~Rasio,
  Phys.\ Rev.\ D {\bf 93}, 084029 (2016)
  [arXiv:1602.02444 [astro-ph.HE]].

\end{thebibliography}
\end{document}